\documentclass{ifacconf}

\usepackage{graphicx}      
\usepackage{natbib}        
\usepackage{amsfonts}
\usepackage{amsmath,amssymb}   
\usepackage{algorithm}
\usepackage{algorithmic}
\usepackage{enumitem}
\usepackage{threeparttable}
\usepackage{multirow}
\usepackage{booktabs}
\usepackage{xcolor}
\usepackage{url}

\usepackage{scalerel}

\newtheorem{remark}{Remark}
\newtheorem{lemma}{Lemma}
 
\newcommand{\diag}{\mathop{\rm diag}\nolimits}

\newcommand{\rr}{{\mathbb R}}
\newcommand{\ba}[1]{\begin{array}{#1}}
\newcommand{\ea}{\end{array}}

\newcommand*{\pdot}{\mathbin{\scalerel*{\boldsymbol\odot}{\circ}}}

\begin{document}
\begin{frontmatter}

\title{Time-Certified and Efficient NMPC via Koopman Operator\thanksref{footnoteinfo}} 

\thanks[footnoteinfo]{ This research was supported by the Ralph O’Connor Sustainable Energy Institute at Johns Hopkins University.}

\author[First]{Liang Wu} 
\author[Second]{Yunhong Che}
\author[Third]{Bo Yang}
\author[Fourth]{Kangyu Lin}
\author[First]{J\'an Drgo\v na}
    
\address[First]{Johns Hopkins University, MD 21218, USA \\
e-mail: \{wliang14, jdrgona1\}@jh.edu.}
\address[Second]{Massachusetts Institute
of Technology, MA 02139, USA \\
e-mail: yunhche@mit.edu.}
\address[Third]{Tsinghua University, Beijing 100084, China \\
e-mail: yang-b21@mails.tsinghua.edu.cn.}
\address[Fourth]{Kyoto University, Graduate School of Informatics, JPN \\
e-mail: k-lin@sys.i.kyoto-u.ac.jp.}
\begin{abstract}
Certifying and accelerating execution times of nonlinear model predictive control (NMPC) implementations are two core requirements. Execution-time certificate guarantees that the NMPC controller returns a solution before the next sampling time, and achieving faster worst-case and average execution times further enables its use in a wider set of applications. However, NMPC produces a nonlinear program (NLP) for which it is challenging to derive its execution time certificates. Our previous works, \citep{wu2025direct,wu2025time} provide \textit{data-independent} execution time certificates (certified number of iterations) for box-constrained quadratic programs (BoxQP). To apply the time-certified BoxQP algorithm \citep{wu2025time} for state-input constrained NMPC, this paper \textit{i)} learns a linear model via Koopman operator; \textit{ii)} proposes a dynamic-relaxation construction approach yields a structured BoxQP rather than a general QP; \textit{iii)} exploits the structure of BoxQP, where the dimension of the linear system solved in each iteration is reduced from $5N(n_u+n_x)$ to $Nn_u$ (where $n_u, n_x, N$ denote the number of inputs, states, and length of prediction horizon), yielding substantial speedups (when $n_x \gg n_u$, as in PDE control).
\end{abstract}
\begin{keyword}
Nonlinear model predictive control, Execution time certificate, Koopman Operator.
\end{keyword}

\end{frontmatter}

\section{Introduction}
Model predictive control (MPC) is a model-based optimal control method widely used in manufacturing, energy systems, and robotics. At each sampling instant, MPC solves an online optimization problem defined by a prediction model, constraints, and an objective. 

Two key requirements for deploying MPC in production are achieving (i) low average execution time, and (ii) a worst-case execution time that remains below the sampling period. While most works focus on improving average execution time \cite{ferreau2014qpoases, stellato2020osqp, wu2023simple, wu2023construction}, the more critical requirement is certifying worst-case execution time, as reflected in the standard assumption that each MPC optimization must be solved before the next sampling instant.

The execution time is computed from the worst-case number of floating-point operations ([flops]) using the approximate relation
$$
\textrm{execution time} = \frac{\textrm{total [flops] required by the algorithm}}{\textrm{average [flops] processed per second}}~[s].
$$
where the denominator depends primarily on the embedded processor technology. Determining the worst-case total [flops] requires knowing the worst-case number of iterations of the optimization algorithm used in the NMPC, provided that each iteration has a known and uniform [flops].

Certifying the number of iterations of an iterative optimization algorithm is an open challenge in online parametric NMPC. Generally, the required number of iterations depends on the convergence speed, and the distance between the optimal point and the initial point. The latter is hard to bound in advance in parametric NMPC scenarios, where the problem data varies with the feedback states at each sampling time. For example, NMPC schemes that reformulate the NLP as a sequence of QPs, via successive linearization or real-time iteration \cite{gros2020linear}, result in a Hessian matrix that varies over time.

\textit{Execution time certificate} problem has recently become an active research topic in MPC field, see \citep{richter2011computational, giselsson2012execution, patrinos2013accelerated, cimini2017exact, arnstrom2019exact, cimini2019complexity,arnstrom2020complexity, arnstrom2021unifying, okawa2021linear, wu2025direct, wu2025quadratic}. Among these works, our prior studies \citep{wu2025direct, wu2025quadratic} provide iteration bounds that are both simple to compute and data-independent. Thus, we applied them for certifying execution times of input-constrained NMPC problems, see \cite{wu2024time, wu2024execution}, where the Koopman framework and RTI scheme are used, respectively. But, this is limited to input-constrained NMPC problems because the execution-time-certified algorithm in \cite{wu2025direct, wu2025quadratic} is only for box-constrained quadratic programs (BoxQP). Later, Ref. \citep{wu2025eiqp} provided execution-time certification and infeasibility detection for general QPs, though its execution time certificate imposes a notable loss in practical speed.

Our latest work \citep{wu2025time} developed a predictor-corrector interior-point method (IPM) based BoxQP algorithm, which preserves execution time certificate and comparable computation efficiency at the same time. Based on that, this paper develops a time-certified and efficient NMPC solution via Koopman operator.

\subsection{Contributions}
To certify execution times for state–input constrained NMPC, this paper makes the following contributions:
\begin{enumerate}
    \item[i)] learns a linear high-dimensional model via Koopman operator; 
    \item[ii)] proposes a dynamic-relaxation construction approach that puts $\ell_2$ norm of the multi-step prediction model into the objective (rather than handled as equality constraints), thereby yielding a structured BoxQP rather than a general QP; 
    \item[iii)] exploits the structure of BoxQP based on the algorithm framework \citep{wu2025time}, where the dimension of the linear system solved in each iteration is reduced from $5(n_u+n_x)N_p$ to $n_u N_p$ (where $n_u,n_x,N_p$ denote the number of inputs, states, and length of prediction hoirzon), yielding substantial speedups (when $n_x \gg n_u$, as in PDE control).
\end{enumerate}

\section{Problem formulations}
 This article considers a nonlinear MPC problem (NMPC) for tracking, as shown in \eqref{eqn_nmpc}, 
 
\rule{\linewidth}{0.8pt}
\textbf{Nonlinear MPC:}
\begin{subequations}\label{eqn_nmpc}
\begin{align}
\min & \sum_{k=1}^{N-1} 
\|x_{k+1}-x_r\|_{W_x}^2 + \|u_{k}-u_r\|_{W_{u}}^2 \label{subeq:nmpc-cost}\\
& \qquad+ \|u_k-u_{k-1}\|_{W_{\Delta u}}^2 \nonumber\\
\text{s.t. }&x_0 = x(t), \; u_{-1}=0, \\[-2pt]
& x_{k+1} = f(x_k,u_k),\quad k=0,\dots,N-1 \label{subeq:nmpc-dyn} \\[-2pt]
&  -\mathbf{1} \leq x_{k+1}  \leq \mathbf{1},\quad k=0,\dots,N-1 \label{subeq:mpc-state-constraints} \\[-2pt]
& -\mathbf{1}\leq u_{k} \leq \mathbf{1},\quad k=0,\dots,N-1
\end{align}
\end{subequations}
\rule{\linewidth}{0.8pt}
where $x(t)$ is the feedback state at the sampling time $t$,  $u_k\in\rr^{n_u}$ and $x_k\in\rr^{n_x}$ denote the control input and state at the $k$th time step, the prediction horizon length is $N$, $x_r$ and $u_r$ denote the desired tracking reference signal for the state and control input, respectively, and $W_x\succ0$, $W_u\succ0$, and $W_{\Delta u}\succ0$ are weighting matrices for the deviation of the state tracking error, control input tracking error, and control input increment, respectively. Assume that $W_x$, $W_u$, and $W_{\Delta u}$ are diagonal. The equality constraint \eqref{subeq:nmpc-dyn} represents the nonlinear discrete-time dynamical system. Without loss of generality, the state and control input constraints, $[x_{\min},x_{\max}]$ and $[u_{\min},u_{\max}]$, are scaled to $[-\mathbf{1},\mathbf{1}]$.

NMPC \eqref{eqn_nmpc} is a nonconvex nonlinear program, which can be infeasible, can be sensitive to the noise-contaminated $x(t)$, can have slow solving speed, and can lack convergence guarantees and execution time certificate. A Koopman framework for MPC \citep{korda2018linear}, which transforms NMPC \eqref{eqn_nmpc} into a convex QP problem via data-driven Koopman approximations, allows the use of execution-time-certified and computationally efficient QP algorithms for real-time applications.

\section{Preliminary: Koopman transforms NMPC into general QP}
The Koopman operator \citep{koopman1931hamiltonian} provides a globally linear representation of nonlinear dynamics. In practice, the infinite-dimensional Koopman operator is truncated and approximated using data-driven Extended Dynamic Mode Decomposition (EDMD) methods \citep{williams2015data,williams2016extending,korda2018convergence}. In EDMD specifically, the set of extended observables is designed as the ``lifted" mapping, $\left[\begin{array}{@{}c@{}}
        \psi(x) \\
        \mathbf{u}(0)
    \end{array}\right]\!,$
where $\mathbf{u}(0)$ denotes the first component of the sequence $\mathbf{u}$ and $\psi(x)\triangleq\left[\psi_1(x), \cdots{}, \psi_{n_\psi}(x)\right]^\top$ ($n_\psi \gg n_x$) is chosen from a basis function, e.g., Radial Basis Functions (RBFs) used in \cite{korda2018linear}, instead of directly solving for them via optimization. In particular, the approximate Koopman operator identification problem is reduced to a least-squares problem, which assumes that the sampled data $\{(
x_j,\mathbf{u}_j),(x_j^+,\mathbf{u}_j^+)\}$ (where $j$ denotes the index of data samples and the superscript $+$ denotes the value at the next time step) are collected with the update mapping $\left[\begin{array}{@{}c@{}}
    x_j^+ \\
    \mathbf{u}_j^+
\end{array}\right]\! =\! \left[\begin{array}{@{}c@{}}
     f(x_j,\mathbf{u}_j(0))  \\
     \boldsymbol{S}\mathbf{u}_j
\end{array}\right]$. Then an approximation of the Koopman operator, $\mathcal{A}\triangleq [A \;\;B]$, can be obtained by solving
\begin{equation}
  J(A,B) = \min_{A,B} \sum_{j=1}^{N_d}\|\psi(x_j^+)-A\psi(x_j)-B\mathbf{u}_j(0)\|_2^2.  
\end{equation}
According to \cite{korda2018linear}, if the designed lifted mapping $\psi(x)$ contains the state $x$ after the re-ordering $\psi(x)\leftarrow [x^\top, \psi(x)]^\top$, then $C=[I,0]$.
The learned linear Koopman predictor model is given as
\begin{equation}\label{eqn_Koopman_linear}
\psi_{k+1} = A \psi_k + Bu_k,~ x_{k+1} = C \psi_{k+1},
\end{equation}
where $\psi_k\triangleq\psi(x_k) \in\rr^{n_\psi}$ denotes the lifted state space and with $\psi_0 = \psi(x(t))$. If~\eqref{eqn_Koopman_linear} has a high lifted dimension (for a good approximation), its use in MPC will not increase the dimension of the resulting QP if the high-dimensional observables $\psi_k$ are eliminated via
\begin{equation}\label{eqn_X_U_E_F}
\mathrm{col}(x_{1},x_{2},\cdots,x_{N})
= \mathbf{E} \psi\big(x(t)\big) + \mathbf{F} \mathrm{col}( u_{0},u_{1},\cdots,u_{N-1}),
\end{equation}
where
\begin{equation}
     \mathbf{E} \triangleq\! \left[\begin{array}{c}
    CA  \\
    CA^2 \\
    \vdots \\
    CA^N
\end{array}\right]\!,\ \  \mathbf{F} \triangleq \left[\begin{array}{cccc}
     CB & 0 & \cdots & 0  \\
     CAB & CB & \cdots & 0 \\
     \vdots & \vdots & \ddots & \vdots \\
     CA^{N-1}B &  CA^{N-2}B & \cdots & CB
\end{array}\right]\!.    
\end{equation}
Then, by embedding \eqref{eqn_X_U_E_F} into the quadratic objective \eqref{subeq:nmpc-cost} and the state constraint \eqref{subeq:mpc-state-constraints}, NMPC \eqref{eqn_nmpc} can be reduced to a compact \textit{general} QP with the decision vector $z\triangleq \mathrm{col}(u_0,\cdots{},u_{N-1})\in\rr^{N\times n_u}$ as shown in \eqref{eqn_KoopmanMPC_QP}, where $\bar{W}_x\triangleq\mathrm{blkdiag}(W_x,\cdots{},W_x)$, $\bar{W}_u\triangleq\mathrm{blkdiag}(W_u,\cdots{},W_u)$, $\bar{R}\triangleq\mathrm{blkdiag}(\bar{W}_{\Delta u},\cdots{},\bar{W}_{\Delta u},\bar{W}_{\Delta u}^N)$ $\Bigg(\footnotesize\bar{W}_{\Delta u}=\left[\begin{array}{cc}
    2W_{\Delta u} &  -W_{\Delta u} \\
    -W_{\Delta u} &  2W_{\Delta u}
\end{array}\right]$ and $\footnotesize\bar{W}_{\Delta u}^N=\left[\begin{array}{cc}
    2W_{\Delta u} &  -W_{\Delta u} \\
    -W_{\Delta u} &  W_{\Delta u}
\end{array}\right]\Bigg)$ and $h \triangleq \mathbf{F}^\top \bar{W}_x(\mathbf{E}\psi(x(t))-\bar{x}_r)-\bar{W}_u\bar{u}_r$ ($\bar{x}_r=\mathrm{repmat}(x_r)$ and $\bar{u}_r=\mathrm{repmat}(u_r)$).
Note that the computation of the high-dimensional observable $\psi(x(t))$ is performed once and will not be involved in the iterations of QP, which minimizes a side effect of the high-dimensional Koopman operator. 

\rule{\linewidth}{0.6pt}
\textbf{NMPC $\rightarrow$ Koopman-QP:}
\begin{subequations}\label{eqn_KoopmanMPC_QP}
\begin{align}
\min_{z} &~ z^\top\!\left(\mathbf{F}^\top \bar{W}_x \mathbf{F}+ \bar{W}_u + \bar{R} \right) \!z + 2z^\top h \label{eqn_KoopmanMPC_QP_obj}\\
\text{s.t.} &~ -\mathbf{1}\leq \mathbf{E}\psi(x_t)+\mathbf{F}z \leq \mathbf{1} \label{eqn__KoopmanMPC_QP_state_constraint} \\
&~ - \mathbf{1} \leq z \leq \mathbf{1}
\end{align}
\end{subequations}  
\rule{\linewidth}{0.6pt}
The resulting Koopman-QP \eqref{eqn_KoopmanMPC_QP} has $N n_u$ decision variables and $2N(n_x+n_u)$ inequality constraints, independent of $n_\psi$. It may be infeasible due to inevitable modeling errors in the data-driven Koopman approximation. By softening the state constraints with slack $\epsilon$ and adding a large penalty term $\rho_\epsilon \epsilon^2$ to the objective, feasibility is guaranteed. The condensed Koopman-QP is a general QP without exploitable structure, and currently there is no structure-tailored QP solver for Koopman-MPC to enable fast real-time computation.

\section{Methodology (1): Koopman-BoxQP via Dynamics-relaxed Construction}
To simultaneously handle potential infeasibility and obtain faster and certified execution time, this article proposes a \textit{dynamics-relaxed} construction that transforms the Koopman-MPC problem into an always-feasible parametric BoxQP. The \textit{dynamic relaxation approach} does not strictly enforce the Koopman prediction model \eqref{eqn_X_U_E_F} but instead incorporates it into the objective through a penalty term, which results in a strongly convex BoxQP  \eqref{eqn_KoopmanMPC_BoxQP}, where $U\triangleq\mathrm{col}(u_0,\cdots{},u_{N-1})$, $X\triangleq\mathrm{col}(x_1,\cdots{},x_{N})$, and $\rho>0$ is a large penalty parameter reflecting the confidence in the Koopman model’s accuracy. 

\rule{\linewidth}{0.8pt}
\textbf{Nonlinear MPC $\rightarrow$ \textit{Koopman-BoxQP}}
{\small
\begin{equation}\label{eqn_KoopmanMPC_BoxQP}
\begin{aligned}
\min_{U,X} &~ (X-\bar{x}_r)^\top \bar{W}_x(X-\bar{x}_r) + (U-\bar{u}_r)^\top \bar{W}_u(U-\bar{u}_r)\\
&~  + U^\top \bar{R} U + \rho \|X- \mathbf{E} \psi\big(x(t)\big) - \mathbf{F}U\|_2^2  \\
\text{s.t.} &~ -\mathbf{1} \leq U \leq \mathbf{1}, \\
&~ -\mathbf{1} \leq X \leq \mathbf{1}
\end{aligned}
\end{equation}
}
\rule{\linewidth}{0.8pt}

For simplicity, we denote the decision vector $z\triangleq\mathrm{col}(U,X)\in\rr^n$ ($n=N (n_u+n_x)$), and the proposed \textit{Koopman-BoxQP} \eqref{eqn_KoopmanMPC_BoxQP} is constructed as shown in \eqref{eqn_Box_QP},
\begin{equation}\label{eqn_Box_QP}
    \begin{aligned}
        \min_{z}&~\frac{1}{2}z^\top H z + z^\top h(x(t))\\
        \text{s.t.}&~ -\mathbf{1}\leq z \leq \mathbf{1}        
    \end{aligned}
\end{equation}
where
\begin{equation}\label{eqn_H_def}
H\triangleq \rho \! \left[\begin{array}{@{}cc@{}}
    \mathbf{F}^\top \mathbf{F} & -\mathbf{F}^\top \\
    -\mathbf{F} & I 
\end{array} \right] + \left[\begin{array}{@{}cc@{}}
    \bar{W}_u+\bar{R}&  \\
     & \bar{W}_x
\end{array}\right]\!\succ0,
\end{equation}
\begin{equation}
 h(x(t))\triangleq \rho\!\left[\begin{array}{@{}c@{}}
      \mathbf{F}^\top\mathbf{E}  \\
      -\mathbf{E} 
 \end{array}\right]\! \psi(x(t)) - \left[\begin{array}{@{}c@{}}
      \bar{W}_u\bar{u}_r  \\
      \bar{W}_x\bar{x}_r 
 \end{array}\right].        
\end{equation}
The dynamics-relaxed \textit{Koopman-BoxQP} \eqref{eqn_KoopmanMPC_BoxQP} can be viewed as an alternative approach to softening the state constraints, since the error in satisfying the prediction model can be equivalently interpreted as a relaxation of the state constraints. 

\section{Methodology (2): Time-certified and Efficient IPM Algorithm for BoxQP}
According to \cite[Ch 5]{boyd2004convex}, the Karush–Kuhn–Tucker (KKT) condition of Box-QP \eqref{eqn_Box_QP} is the following nonlinear equations,
\begin{subequations}\label{eqn_KKT}
\begin{align}
    Hz + h(x(t)) + \gamma - \theta = 0,\label{eqn_KKT_a}\\
    z + \phi - \mathbf{1}_n=0,\label{eqn_KKT_b}\\
    z - \psi + \mathbf{1}_n=0,\label{eqn_KKT_c}\\
    (\gamma,\theta,\phi,\psi)\geq0,\label{eqn_KKT_d}\\
    \gamma \pdot \phi = 0,\label{eqn_KKT_e}\\
    \theta \pdot \psi = 0,\label{eqn_KKT_f}
\end{align}
\end{subequations}
where $\gamma,\theta$ are the Lagrangian variables of the lower and upper bound, respectively, and $\phi,\psi$ are the slack variables of the lower and upper bound, respectively. $\pdot$ represents the Hadamard product, i.e., $\gamma\pdot \phi = \mathrm{col}(\gamma_1\phi_1,\gamma_2\phi_2,\cdots{},\gamma_n\phi_n)$.

Path-following primal–dual IPMs are categorized into two types: \textit{feasible} and \textit{infeasible}, distinguished by whether the initial point satisfies Eqns. \eqref{eqn_KKT_a}–\eqref{eqn_KKT_d}. For the complementarity constraints \eqref{eqn_KKT_e}–\eqref{eqn_KKT_f}, \textit{feasible} path-following IPMs require the initial point to lie in a narrow neighborhood. To demonstrate this, let us denote the feasible region by $\mathcal{F}$, i.e.,
\begin{equation}
     \mathcal{F}=\{(z,\gamma,\theta,\phi,\psi):\eqref{eqn_KKT_a}\mathrm{-}\eqref{eqn_KKT_c},(\gamma,\theta,\phi,\psi)\geq0\}
\end{equation}
and the set of strictly feasible points by
\begin{equation}
    \mathcal{F}^+\triangleq\{(z,\gamma,\theta,\phi,\psi):\eqref{eqn_KKT_a}\mathrm{-}\eqref{eqn_KKT_c},(\gamma,\theta,\phi,\psi)>0\}
\end{equation}
We also consider the neighborhood
\begin{equation}
    \mathcal{N}(\beta)\triangleq\left\{(z,\gamma,\theta,\phi,\psi)\in\mathcal{F}^+: \left\|\left[\begin{array}{c}
         \gamma\pdot \phi  \\
         \theta\pdot \psi 
    \end{array}\right]-\mu \mathbf{1}_{2n} \right\|_2\leq\beta\mu\right\}
\end{equation}
where the duality measure $\mu\triangleq\frac{\gamma^\top \phi+\theta^\top\psi}{2n}$ and $\beta\in[0,1]$. \textit{Feasible} path-following IPMs require the initial point:
\begin{equation}\label{eqn_initial_point_requirement}
    (z^0,\gamma^0,\theta^0,\phi^0,\psi^0)\in\mathcal{N}(\beta),
\end{equation}
and computing such a point is typically expensive for general strictly convex QPs.

 \subsection{Cost-free initialization for Feasible IPMs}
Inspired by our previous work \cite{wu2025direct}, which first showed that Box-QP admits cost-free initialization for feasible IPMs, we propose the following initialization to ensure $(z^0,\gamma^0,\theta^0,\phi^0,\psi^0)\in\mathcal{N}(\beta)$.
\begin{remark}\label{remark_initialization}
For $h(x(t))=0$, the optimal solution of Box-QP \eqref{eqn_Box_QP} is $z^*=0$. For $h\neq0$,  first scale the objective as 
\[
\min_z \tfrac{1}{2} z^\top (2\lambda H) z + z^\top (2\lambda h(x(t)))
\]
which does not affect the optimal solution 
and can ensure the initial point lies in $\mathcal{N}(\beta)$ if $\lambda\leftarrow\frac{\beta}{\sqrt{2}\|h\|_2}$. Then \eqref{eqn_KKT_a} is replaced by 
\[
2\lambda H z+2\lambda h(x(t))+\gamma-\theta=0
\]
the initialization strategy for Box-QP \eqref{eqn_Box_QP} 
\begin{equation}\label{eqn_initialization_stragegy}
\begin{aligned}
&z^0 = 0,~\gamma^0 =\mathbf{1}_n - \lambda h(x(t)),~\theta^0 =\mathbf{1}_n + \lambda h(x(t)),\\
&~\phi^0 = \mathbf{1}_n, ~\psi^0 = \mathbf{1}_n,       
\end{aligned}
\end{equation}
which clearly places this initial point in $\mathcal{N}(\beta)$ by its definition in Eqn. \eqref{eqn_initial_point_requirement} (for example, $\left\|\left[\begin{array}{c}
         \gamma^0\pdot \phi^0  \\
         \theta^0\pdot \psi^0 
    \end{array}\right]-\mu \mathbf{1}_{2n} \right\|_2=\beta\mu$, where $\mu=1$). In particular, this letter chooses $\beta=\frac{1}{4}$, then $\lambda=\frac{1}{4\sqrt{2}\|h(x(t))\|_2}$.
\end{remark}

\subsection{Algorithm and worst-case iteration complexity}
For simplicity, we introduce
\[
v\triangleq\mathrm{col}(\gamma,\theta)\in\rr^{2n},~ s\triangleq\mathrm{col}(\phi,\psi)\in\rr^{2n}.
\]
According to Remark \ref{remark_initialization}, we have $(z,v,s)\in\mathcal{N}(\beta)$. Then, all the search directions
$(\Delta z,\Delta v,\Delta s)$
(for both predictor and corrector steps) are obtained as solutions of the following system of linear equations:
\begin{subequations}\label{eqn_Newton}
    \begin{align}
    (2\lambda H)\Delta z+ \Omega \Delta v &=0\label{eqn_Newton_a}\\
    \Omega^\top \Delta z + \Delta s & =0\label{eqn_Newton_b}\\
    s\pdot \Delta v + v\pdot \Delta s &=  \sigma \mu \mathbf{1}_{2n} - v\pdot s\label{eqn_Newton_c}
    \end{align}
\end{subequations}
where $\Omega=[I,-I] \in\mathbb{R}^{n \times 2n}$, $\sigma$ is chosen $0$ in predictor steps and $1$ in correctors steps, respectively, and $\mu\triangleq\frac{v^\top s}{2n}$ denotes the duality measure.
\begin{remark}\label{remark_positive}
Eqns. \eqref{eqn_Newton_a} and \eqref{eqn_Newton_b} imply that
\[
\Delta v^{\!\top\!}
\Delta s=\Delta v^{\!\top\!} (-\Omega^{\!\top\!}\Delta z)=\Delta z^{\!\top\!} (2\lambda H)\Delta z\geq0,
\]
which is critical in the following iteration complexity analysis.
\end{remark}
By letting
\begin{equation}\label{eqn_Delta_gamma_theta_phi_psi}
    \begin{aligned}
        &\Delta \gamma= \sigma\mu\frac{1}{\phi}- \gamma + \frac{\gamma}{\phi}\Delta z,~\Delta \theta= \sigma\mu\frac{1}{\psi}-\theta-\frac{\theta}{\psi}\Delta z,\\
        &\Delta\phi = - \Delta z,~\Delta\psi = \Delta z,    
    \end{aligned}
\end{equation}
Eqn. \eqref{eqn_Newton} can be reduced into a more compact system of linear equations,
\begin{equation}\label{eqn_compact_linsys}
 \Big(2\lambda H+\diag\!\Big(\frac{\gamma}{\phi}\Big) + \diag\!\Big(\frac{\theta}{\psi}\Big) \!\Big) \Delta z= \sigma\mu\left(\frac{1}{\phi} - \frac{1}{\psi}\right) + \gamma - \theta.   
\end{equation}
The proposed feasible adaptive-step predictor-corrector IPM algorithm for Box-QP \eqref{eqn_Box_QP} is first described in Algorithm \ref{alg_PC_IPM}. 
\begin{algorithm}
    \caption{Time-certified predictor-corrector IPM for Box-QP \eqref{eqn_Box_QP}} \label{alg_PC_IPM}
    \textbf{Input}: Given a strictly feasible initial point  $(z^0,v^0,s^0)\in \mathcal{N}(1/4)$ from Remark \ref{remark_initialization} and a desired optimal level $\epsilon$. Then the worst-case iteration bound is $N_{\max}=\left\lceil \frac{\log(\frac{2n}{\epsilon})}{-2\log\left(1-\frac{0.2348}{\sqrt{2n}}\right)}\right\rceil$.
    \vspace*{.1cm}\hrule\vspace*{.1cm}

    \textbf{for} $k=0,1, 2,\cdots{},N_{\max}-1$ \textbf{do}
    \begin{enumerate}[label*=\arabic*., ref=\theenumi{}]
        \item if $(v^k)^\top s^k\leq \epsilon$, then break;
        \item Compute the predictor direction $(\Delta z_p,\Delta v_p,\Delta s_p)$ by solving Eqn. \eqref{eqn_Newton} with $(z,s,v)=(z^k,v^k,s^k)$, $\sigma\leftarrow0$, and $\mu\leftarrow\mu^k=\frac{(v^k)^\top s^k}{2n}$ (involving Eqns. \eqref{eqn_compact_linsys} and \eqref{eqn_Delta_gamma_theta_phi_psi});
        \item $\Delta \mu_{p}\leftarrow\frac{(\Delta v_p)^\top \Delta s_p}{2n}$;
        \item $\alpha^k\leftarrow \min\left(\frac{1}{2},\sqrt{\frac{\mu^{k}}{8\|\Delta v_p \pdot \Delta s_p -\Delta \mu_{p}\mathbf{1}_{2n}\|}}\right)$;
        \item $\hat{z}^k\leftarrow z^k+\alpha^k\Delta z_p,~\hat{v}^k\leftarrow v^k+\alpha^k\Delta v_p,~\hat{s}^k\leftarrow s^k+\alpha^k\Delta s_p$;
        \item Compute the corrector direction $(\Delta z_c,\Delta v_c,\Delta s_c)$ by solving Eqn. \eqref{eqn_Newton} with $(z,v,s)=(\hat{z}^k,\hat{v}^k,\hat{s}^k)$, $\sigma\leftarrow1$, and $\mu\leftarrow\hat{\mu}^k=\frac{(\hat{v}^k)^\top \hat{s}^k}{2n}$ (involving Eqns. \eqref{eqn_compact_linsys} and \eqref{eqn_Delta_gamma_theta_phi_psi});
        \item $z^{k+1}\leftarrow \hat{z}^k+\Delta z_c,~v^{k+1}\leftarrow \hat{v}^k+\Delta v_c,~s^{k+1}\leftarrow \hat{s}^k+\Delta s_c$;
    \end{enumerate}
    \textbf{end}\vspace*{.1cm}\hrule\vspace*{.1cm}
    \textbf{Output:} $z^{k+1}$.
\end{algorithm}

The convergence analysis and worst-case iteration complexity of Algorithm \ref{alg_PC_IPM} follow directly from our earlier work \cite{wu2025time}.
\begin{lemma}\label{lemma_1}
(see \cite[Thm. 1]{wu2025time})
       Let $\{(z^k,v^k,s^k) \}$ be generated by Algorithm \ref{alg_PC_IPM}. Then
    \begin{equation}
        \mu^{k+1}\leq\left(1-\frac{0.2348}{\sqrt{2n}}\right)^2\mu^k
    \end{equation}

    Furthermore, Algorithm \ref{alg_PC_IPM} requires at most
    \begin{equation}\label{eqn_N_max}
        N_{\max}=\left\lceil \frac{\log(\frac{2n}{\epsilon})}{-2\log\left(1-\frac{0.2348}{\sqrt{2n}}\right)}\right\rceil.
    \end{equation} 
\end{lemma}
Note that in practice, Algorithm \ref{alg_PC_IPM} exhibits $O(n^{0.25})$ or $O(\log n)$-order iteration complexity due to the conservativeness of our proof, see \cite{wu2025time}.

\subsection{Efficient computation of Newton systems}
At each iteration of Algorithm \ref{alg_PC_IPM}, Steps 2 and 6 of Algorithm \ref{alg_PC_IPM} compute the predictor and corrector directions, respectively, using the same coefficient matrix as given by (according to Eqn. \eqref{eqn_H_def}):
\[
2\lambda H+\diag\!\left(\frac{\gamma^k}{\phi^k}+\frac{\theta^k}{\psi^k}\right) = \left[\begin{array}{@{}cc@{}}
        \bar{H}_{11} &  -2\lambda\rho \mathbf{F}^\top\\
        -2\lambda\rho \mathbf{F} & \bar{H}_{22}
    \end{array}\right]\!\succ0
\]
with
\[
\begin{aligned}
\bar{H}_{11}&\triangleq2\lambda\rho\mathbf{F}^\top\mathbf{F}+2\lambda\bar{W}_u+2\lambda\bar{R}+\diag\!\left(\frac{\gamma_{1:n_1}^k}{\phi_{1:n_1}^k}+\frac{\theta_{1:n_1}^k}{\psi_{1:n_1}^k}\right),\\
    \bar{H}_{22}&\triangleq 2\lambda\rho I +2\lambda\bar{W}_x+\diag\!\left(\frac{\gamma_{n_1+1:n}^k}{\phi_{n_1+1:n}^k}+\frac{\theta_{n_1+1:n}^k}{\psi_{n_1+1:n}^k}\right),
\end{aligned}
\]
where $\bar{H}_{11}\succ0\in\rr^{N n_u\times N n_u}$, $\bar{H}_{22}\succ0\in\rr^{ n_x\times  n_x}$. By assumption the weighting matrix $W_x$ is diagonal, thus $\bar{W}_x$ and $\bar{H}_{22}$ are also diagonal.

\begin{remark}
By exploiting the diagonal structure of $\bar{H}_{22}$ and the Schur Complement, the decomposition cost can be reduced to the  element-wise division for and the  Cholesky factorization 
\[
\begin{aligned}
     \bar{H}_{22}^{-1} &\text{ with the cost } O(Nn_x)\\
     \mathrm{Chol}\left(\bar{H}_{11}-\rho^2 \mathbf{F}^\top \bar{H}_{22}^{-1}\mathbf{F}\right) &\text{ with the cost } O((Nn_u)^3).
\end{aligned}
\]
In summary, the transformation from \eqref{eqn_Newton} to \eqref{eqn_compact_linsys} reduces the problem dimension from $5N(n_u+n_x)$ to $N(n_u+n_x)$, and the subsequent $2\times 2$ block reduction of $H$ in \eqref{eqn_H_def} further reduces it to $N n_u$. 
\end{remark}

\begin{remark}
    In most PDE control applications, the number of discretized spatial states is significantly larger than the number of control inputs, i.e., $n_x\gg n_u$. This is why our proposed Algorithm is not only time-certified but also computationally efficient in large-scale PDE-MPC applications, as demonstrated in Section \ref{sec_numerical_example}.
\end{remark}


\section{Numerical Examples}\label{sec_numerical_example}
This section applies our proposed \textit{Koopman-BoxQP} to solve a PDE-MPC problem with state and control input constraints, which is a large-sized QP problem with $1040$ variables and $2080$ constraints. The PDE plant under consideration is the nonlinear Korteweg-de Vries (KdV) equation that models the propagation of acoustic waves in plasma or shallow water waves \citep{miura1976korteweg} as
\begin{equation}\label{eqn_KdV}
\frac{\partial y(t, x)}{\partial t}+y(t, x) \frac{\partial y(t, x)}{\partial x}+\frac{\partial^3 y(t, x)}{\partial x^3}=u(t,x)
\end{equation}
where $x\in[-\pi,\pi]$ is the spatial variable. Consider the control input $u(t,x)=\sum_{i=1}^4u_i(t)v_i(x)$, in which the four coefficients $\{u_i(t)\}$ are subject to the constraint $[-1,1]$, and $v_i(x)$ are predetermined spatial profiles given as $v_i(x)=e^{-25(x-m_i)^2}$, with $m_1=-\pi/2$, $m_2=-\pi/6$, $m_3=\pi/6$, and  $m_4=\pi/2$. The control objective is to adjust ${u_i(t)}$ so that the spatial profile $y(t,x)$ tracks the given reference signal. The spatial profile $y(t,x)$ is uniformly discretized into $100$ spatial nodes. These 100 discretized nodes serve as the system states, each constrained within $[-1,1]$, while the control inputs are likewise bounded within $[-1,1]$. 

With a prediction horizon of $N=10$, the resulting MPC problem has $1040$ variables and $2080$ constraints. The cost matrices are chosen as $W_x = I_{100}$ and $W_u = 0.05 I_{4}$. The state reference $x_r \in \mathbb{R}^{100}$ is a sinusoidal signal over a $50$ s simulation, while the input reference is constant with $u_r = 0$. Data generation and closed-loop MPC simulation are performed using a Fourier-based spectral method with a split-step scheme for the nonlinear KdV \eqref{eqn_KdV}, with a sampling time of $\Delta t = 0.01$. The closed-loop simulation is configured as follows: (i) Data generation: 1000 trajectories of length 200 are generated from random linear combinations of four spatial profiles with control inputs in $[-1,1]$. (ii) Koopman predictor: The lift $\psi$ includes 100 physical states and 200 thin-plate RBFs ($N_\psi=300$). The matrices $A\in\mathbb{R}^{300\times300}$ and $B\in\mathbb{R}^{300\times4}$ are identified via the Moore–Penrose pseudoinverse, with $C=[I_{100},0]$. (iii) Koopman-BoxQP formulation: Using $(A,B,C)$, the multi-step model is embedded into \eqref{eqn_KoopmanMPC_BoxQP} with $\rho=10^2$, producing a Box-QP with 1040 variables and 2080 constraints.

Before applying Algorithm \ref{alg_PC_IPM} to solve the resulting large-size \textit{Koopman-BoxQP} problem \eqref{eqn_KoopmanMPC_BoxQP} in the closed-loop simulation, we can know in advance that the maximum number of iterations for $n=1040,\epsilon=10^{-6}$ by Lemma \ref{lemma_1}:
\[
N_{\max}=\left\lceil \frac{\log(\frac{2\times 1040}{10^{-6}})}{-2\log\left(1-\frac{0.2348}{\sqrt{2\times 1040}}\right)}\right\rceil=2079
\]
which thus offer a worst-case execution time certificate. 

In practice, Algorithm \ref{alg_PC_IPM} requires an average of $72$ iterations and at most $76$ iterations across all sampling instants during the closed-loop simulation. The execution times of Algorithm \ref{alg_PC_IPM} and other state-of-the-art QP solvers are reported in Tab. \ref{tab_execution_time}, which shows that Algorithm \ref{alg_PC_IPM} is most efficient.

\begin{table}[!htbp]
\caption{Execution time comparison between Algorithm \ref{alg_PC_IPM} and other state-of-the-art QP solvers.}\label{tab_execution_time}
\centering
\begin{tabular}{cc}
\toprule
QP Solver  & Worst-case Execution Time [s]\footnotemark \\ 
\midrule
Quadprog &0.1127 \\
OSQP & $9.7\times 10^{-3}$\\
SCS & $5.8\times10^{-3}$\\
Algorithm \ref{alg_PC_IPM} & $\mathbf{2.163\times10^{-3}}$ \\ 
\bottomrule
\end{tabular}
\end{table}
\footnotetext{The execution time results are based on MATLAB's Quadprog solver, OSQP's and SCS's MATLAB interface (\url{https://osqp.org/} and \url{https://www.cvxgrp.org/scs/}), and C-MEX implementation of Algorithm \ref{alg_PC_IPM}, running on a Mac mini with an Apple M4 Chip (10-core CPU and 16 GB RAM).} 

\begin{figure*}[!htbp]\label{fig}
\begin{picture}(140,110)
\put(0,0){\includegraphics[width=65mm]{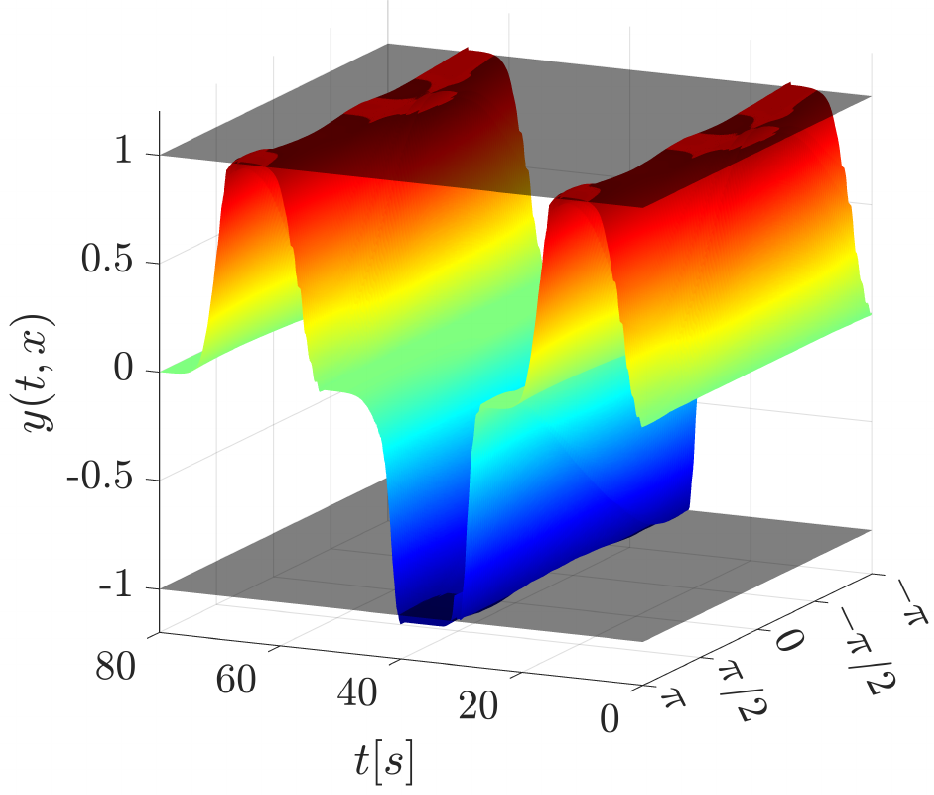}}
\put(190,0){\includegraphics[width=55mm]{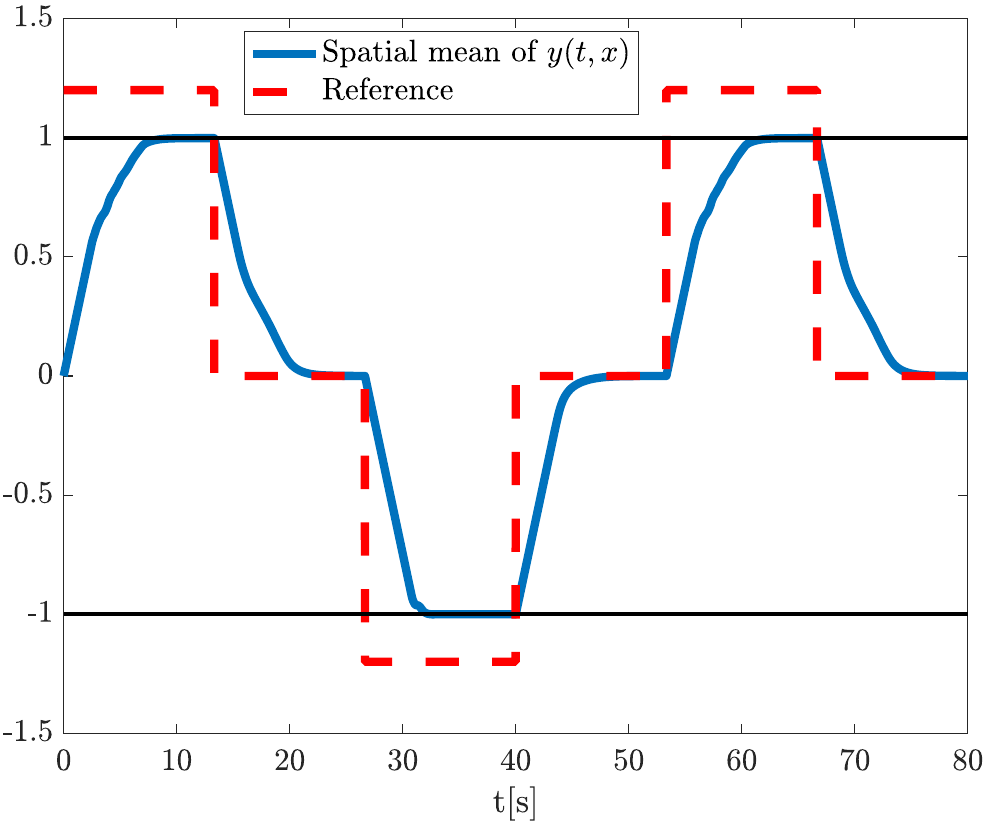}}
\put(360,0){\includegraphics[width=55mm]{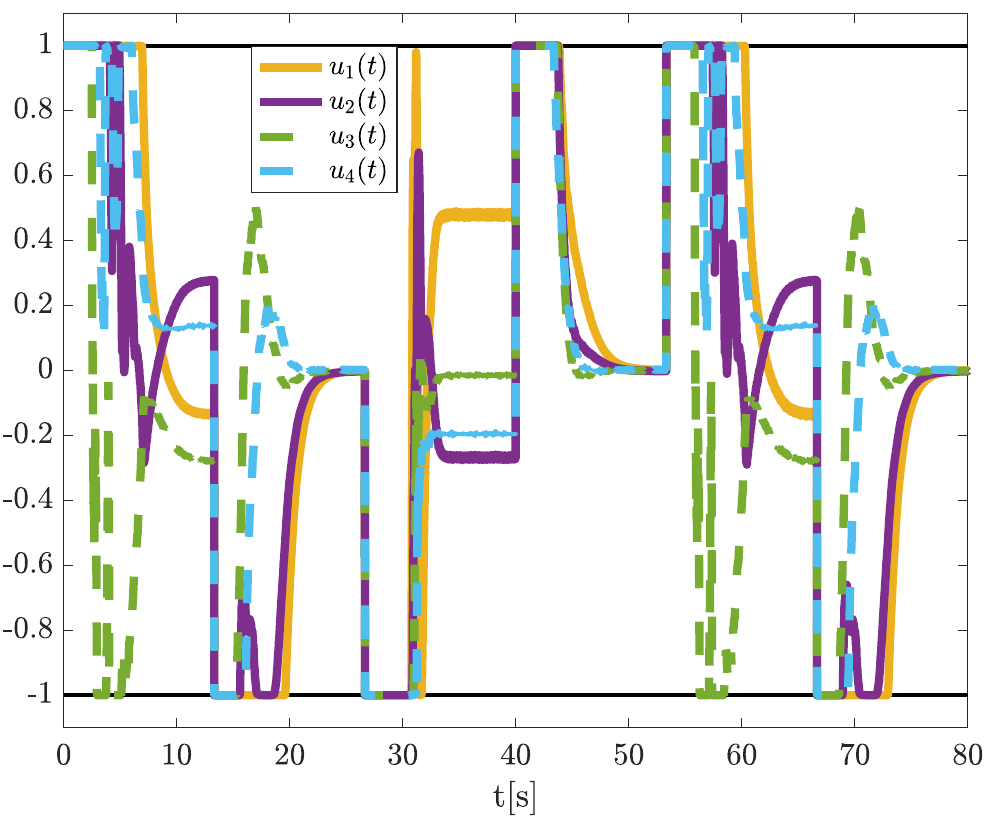}}
\end{picture}
\caption{Closed-loop simulation of the nonlinear KdV system with the dynamics-relaxed \textit{Koopman-BoxQP} controller tracking a time-varying spatial profile reference. Left: time evolution of the spatial profile $y(t,x)$ and the state constraints $[-1,1]$. Middle: spatial mean of the $y(t,x)$ and the state constraints $[-1,1]$. Right: the four control inputs and the control input constraints $[-1,1]$.}
\label{fig_kdv}
\end{figure*}

The closed-loop results in Fig.~\ref{fig_kdv} show that $y(t,x)$ accurately tracks the reference while satisfying both state and input bounds in $[-1,1]$.

\section{Conclusion}
This paper presents a time-certified and efficient NMPC framework based on a dynamic-relaxed \textit{Koopman-BoxQP} formulation that transforms NMPC into a structured Box-QP. Building on the time-certified Box-QP solver in \cite{wu2025time}, we introduce a structure-exploiting linear-algebra technique that reduces the dimension of the resulting linear system, enabling significant speedups. Future work will study the stability and robustness of the dynamics-relaxed \textit{Koopman-BoxQP} framework and compare it against other state-of-the-art optimization solvers.

\section*{DECLARATION OF GENERATIVE AI AND AI-ASSISTED TECHNOLOGIES IN THE WRITING PROCESS}
During the preparation of this work, the authors used ChatGPT to polish paragraphs. After using this tool/service, the authors reviewed and edited the content as needed and take full responsibility for the content of the publication.

\bibliography{ifacconf}             

\end{document}